\documentclass[a4paper,11pt]{article}
\usepackage{pos}
\usepackage{physics}
\usepackage{slashed}
\usepackage{extarrows}
\usepackage{soul}

\title{Extracting the Pion Distribution Amplitude from Lattice QCD through Pseudo-Distributions}
\ShortTitle{Pion DA from Pseudo-Distributions}

\author*[a]{Daniel Kovner}
\author[a,b]{Joe Karpie}
\author[a,b]{Konstantinos Orginos}
\author[b,c]{Anatoly Radyushkin}
\author[d]{Savvas Zafeiropoulos}

\onbehalf{for the HadStruc collaboration}

\affiliation[a]{Department of Physics, William \& Mary,\\
  300 Ukrop Way, Williamsburg, VA 23185}

\affiliation[b]{Thomas Jefferson National Accelerator Facility\\
12000 Jefferson Ave, Newport News, VA 23606}

\affiliation[c]{Physics Department, Old Dominion University, Norfolk, VA 23529, USA}

\affiliation[d]{Aix Marseille Univ, Université de Toulon, CNRS, CPT, Marseille, France}

\emailAdd{dkovner@wm.edu}
\emailAdd{jkarpie@jlab.org}
\emailAdd{kostas@wm.edu}
\emailAdd{radyush@jlab.org}
\emailAdd{savvas.zafeiropoulos@cpt.univ-mrs.fr}
\abstract{The Light-Cone Distribution Amplitude (LCDA) encodes the non-perturbative information of the leading Fock component of the hadron wave function, therefore required for processes including exclusive hadron production. As the pseudo-Nambu-Goldstone boson of QCD, the nonperturbative structure of the pion is of particular interest. Progress on the Lattice QCD calculation of the pion LCDA on ${\cal O}(a)$-improved Wilson fermion ensembles at several lattice spacings is presented. Excited-state systematics are taken into account within a Bayesian Model Averaging framework. A Renormalization-Group-Invariant (RGI) ratio of matrix elements is formed for further extraction of the pion LCDA.
}

\FullConference{The 40th International Symposium on Lattice Field Theory (Lattice 2023)\\
July 31st - August 4th, 2023\\
Fermi National Accelerator Laboratory\\}

\begin{document}
JLAB-THY-24-3982

\maketitle

\section{Introduction}

\paragraph{}
Mapping out the structure of hadrons is crucial to delineating the fundamental properties of QCD and those arising from other Standard Model interactions. In particular, information about hadronic structure, necessary
to describe Deeply Virtual Exclusive Processes is contained in the Light-Cone Distribution Amplitudes (LCDAs or simply DAs), 
introduced originally for the quark-antiquark Fock component of the pion ~\cite{Radyushkin:1977gp, Efremov:1979qk,Lepage:1979zb},  
and  extended~\cite{Lepage:1980fj,Radyushkin:1983wh} 
onto higher Fock states inside a fast-moving hadron. LCDAs encode the distribution of momentum of the hadron among the Fock state degrees of freedom. A systematically controlled non-perturbative computation of the quark-antiquark LCDA in the pion is a necessary ingredient in the description of Deeply Virtual Exclusive Meson Production
cross sections, which require constraints on meson DAs to study proton spin structure through Generalized Parton Distributions.

\paragraph{}
Due to the Euclidean metric signature of Lattice QCD, a naively direct computation of the DA is not possible. Historically, lattice QCD computations of the pion DA and related quantities have been limited to Mellin-moment reconstruction. Only the first couple moments can be computed from local operators on the lattice due to the reduction of continuous $O(3)$ rotational symmetry to the hypercubic group $H_4$~\cite{Beccarini:1995iv}. This symmetry-breaking mixes the moments under renormalization with power-divergent coefficients. With the development of the quasi-distribution~\cite{Ji:2013dva,Ji:2014gla,Ji:2015jwa} and pseudo-distribution~\cite{Radyushkin:2017cyf,Orginos:2017wcl,Joo:2019jct,Joo:2019bzr,Joo:2020spy} formalisms, it is now possible to extract the full distribution. The results to be presented use the Pseudo-Distribution approach. Previous results of the LCDA employing the quasi-distribution approach can be found in~\cite{Zhang:2017bzy,Gao:2022vyh,LatticeParton:2022zqc} or in~\cite{Bali:2018spj} where instead a product of space-like separated currents is considered. In addition, results on the pseudo-DA of the $\eta_c$ meson were also presented in~\cite{SanJosePerez:2023qqx} during this conference.

\section{Pseudo-Distributions}

The pseudo-distribution approach begins with finding a general matrix element that generates the quantity of interest in the light-cone kinematics. In the case of the leading-twist pion DA, the matrix element is
\begin{align}
    {M}^\alpha(p,z) = \mel{\Omega}{\overline{\psi}(-z/2)\gamma^5\gamma^\alpha U[-z/2,z/2]\psi(z/2)}{\pi(p)}
   \label{eq:matelem}
\end{align}
with Lorentz decomposition
\begin{align}
    M^\alpha(p,z) = 2p^\alpha \mathcal{M}(\nu,-z^2) + z^\alpha\mathcal{N}(\nu,-z^2)
    \label{eq:Lorentz}
\end{align}
where $U[a,b]$ denotes a Wilson line from spacetime coordinates $a$ to $b$, $(\mathcal{M},\mathcal{N})$ are functions of Lorentz invariant "Ioffe time" $\nu=-p\cdot z$  and separation $z^2$. The matrix element definition of the LCDA
is exactly recovered by placing the kinematics on the light cone, i.e., $\alpha = +$, $z = z_-$, $p=p_+$. In this frame, only the $\mathcal{M}$ term survives, implying that only knowledge of $\mathcal{M}$ is required to define the LCDA at scale $\mu^2=-1/z^2$, at leading order $O(\alpha_S^0)$,
\begin{align}
    \mathcal{M}(\nu,-z^2) = \int_0^1 \varphi(x,\mu^2=-1/z^2) e^{i\nu(x-1/2)}\, dx.
\end{align}
The isolation of "Ioffe-time distribution"  (ITD) $\mathcal{M}$ is also clear in the equal-time frame with the gauge link along the $z$ direction, $\alpha=t$, $z=(0,0,0,z_3)$, $p=(p_0,0,0,p_3)$. The multiplicative UV divergence~\cite{Ishikawa:2017faj} accompanying the space-like gauge-link is removed by constructing the reduced ITD,
\begin{align}
    \mathfrak{M}(\nu,z^2) = \frac{\mathcal{M}(\nu,-z^2)}{\mathcal{M}(0,-z^2)}.
    \label{eq:ratio}
\end{align}
For sufficiently small $z^2$, perturbative evolution dominates the $z^2$ dependence of $\mathfrak{M}$ as $\log(z^2)$. The $\overline{\rm MS}$ LCDA is then inferred through perturbative matching relations in this regime to safely facilitate the $z^2\to 0$ limit.

As pointed out in ~\cite{Constantinou:2017sej}, one must be careful about the constant operator-mixing between non-local operators containing different Dirac structures with mixing pattern $\Gamma\leftrightarrow\anticommutator{\Gamma}{\slashed{z}}$. In the DA case, the mixing pattern is $\gamma^5\gamma^\alpha \leftrightarrow z^\beta\gamma^5\sigma^{\alpha\beta}$.  By examining Lorentz decomposition of the matrix element of the $z^\beta\gamma^5\sigma^{\alpha\beta}$ operator,
\begin{align}
    M^{\alpha,{\rm mix}}(p,z) = z^\beta (p^\alpha z^\beta-p^\beta z^\alpha)\mathcal{M}^{\rm mix}(\nu,-z^2) = p^\alpha[z^2\mathcal{M}^{\rm mix}(\nu,-z^2)]+z^\alpha[\nu \mathcal{M}^{\rm mix}(\nu,-z^2)]\,,
\end{align}
we find that the only surviving term for the previously mentioned kinematics is suppressed by $z^2$. Since this mixing is not power divergent, and the anomalous dimensions of both operators are the same, the bare matrix element of Eq.~\ref{eq:matelem} admits the Lorentz decomposition of Eq.~\ref{eq:Lorentz} up to an overall multiplicative factor that is UV divergent. 
Therefore, the reduced ITD in Eq.~\ref{eq:ratio} constructed from the  Lorentz-invariant form factor $\mathcal M$, which can be extracted from any appropriately chosen bare matrix element $M^a$, has no UV divergences. We opt to use the $\alpha=t$ Dirac structure which up to $\order{z^2}$ corrections results in the desired reduced ITD. 

Another approach  is to use the $\alpha=z$ Dirac structure, which has no mixing ~\cite{Gao:2022vyh}. However, one still has to deal with an additional source of off-light-cone contamination originating from the Lorentz decomposition,
\begin{align}
    M^{\alpha=z}(p,z) = 2p_z\mathcal{M}(\nu,-z^2) + z_3\mathcal{N}(\nu,-z^2).
\end{align}
This choice of kinematics does not eliminate the second term and therefore does not isolate the desired Lorentz invariant $\mathcal M$. 
Removing UV divergences through ratios with this Dirac structure cannot be achieved by utilizing the rest frame matrix element, due to the $p_z$ coefficient of $\mathcal{M}$ in the Lorentz decomposition. One is then left with forming reduced ITDs for moving frame matrix elements, such as the ratio
\begin{align}
   \overline{M}(\nu,\nu',-z^2) = \frac{ M^{\alpha=z}(p,z)}{ M^{\alpha=z}(p',z)} = \frac{\mathcal{M}(\nu,-z^2)+\frac{z^2}{\nu}\mathcal{N}(\nu,-z^2)}{\mathcal{M}(\nu',-z^2)+\frac{z^2}{\nu'}\mathcal{N}(\nu',-z^2)}.
\end{align}
This choice of kinematics unnecessarily complicates the extraction of the light-cone distribution amplitude. For that reason, we have decided not to use it.

\section{Matrix Element Extraction}

We perform the lattice calculation on a subset of CLS ensembles with $\order{a}$ improved Wilson-fermion actions(~\cite{Fritzsch:2012wq} and Tab.~\ref{tab:lat}), generating the following two-point functions:
\begin{align}
    C^{ij}_{p}(t) &= \expval{O_i(-p,t)\overline{O}_j(p,0)},\quad C^{\Gamma j}_{p,z}(t)=\expval{O_\Gamma(-p,z,t)\overline{O}_j(p,0)},\nonumber\\
    O_i(p,t)&=\sum_{x}e^{ip\cdot x}\overline{\psi}(x)\gamma^{[i]}\psi(x),\quad O_\Gamma(p,z,t) = \sum_{x}e^{ip\cdot x}\overline{\psi}(x)\Gamma U[x,x+z]\psi(x+z),\nonumber\\
    \gamma^{[i]}&\in\{\gamma^5,\gamma^5\gamma^t\},\quad \Gamma = \gamma^5\gamma^t.
\end{align}

\begin{table*}[ht!] 
\centering
{\renewcommand{\arraystretch}{1.2}
\begin{tabular}{ l | c c | c | c c | c}
ID & ~$a$(fm)~ & ~$M_\pi$(MeV) & $L^3 \times T$ & $N_{\rm cfg}$ & $N_{\rm sources}$ & $\zeta$\\\hline\hline
$\widetilde{A5}$ & 0.0749(8) & 446(1) & $32^3 \times 64$ & 1904 & 8 & 0,2,4\\\hline
E5 & 0.0652(6) & 440(5) & $32^3 \times 64$ & 999 & 128 & 0,3,4.5,6\\\hline
N5 & 0.0483(4) & 443(4) & $48^3 \times 96$ & 477 & 32 & 0, 3, 4.5,6\\\hline \hline\end{tabular}
}

\caption{\label{tab:lat}\footnotesize Parameters for the lattices generated by the CLS collaboration using two flavors of $\mathcal{O}(a)$ improved Wilson fermions. More details about these ensembles can be found in~\cite{Fritzsch:2012wq}. 
}
\end{table*}

Momentum-Gaussian smeared interpolators are employed to enhance the signal of large momentum-projected correlators ~\cite{Bali:2016lva}. The smearing parameters (Tab. ~\ref{tab:lat}) are selected to amortize the computational cost, signal improvement, and excited state contamination across accessible lattice momenta. The $CP$ properties of the pion are exploited to double the number of separations by first generating two-point functions with $O_{\Gamma}(p,z,t)$, and then scaling the result by $\exp(-i\nu/2)$. The consequent DA matrix element has its gauge link centered at the origin. The non-local DA matrix element is extracted from the ratio of two-point functions
\begin{align}
    R^{\Gamma,i}_{p,z}(t) = e^{-i\nu/2}\frac{C^{\Gamma i}_{p,z}(t)}{C^{\Gamma i}_{p,z=0}(t)} \stackrel{t>>0}{\approx} \frac{M^\Gamma(p,z)}{E(p)f_\pi} + \order{e^{-\Delta E(p) t}}. \label{eq:ratio}
\end{align}
The large time extrapolation is performed through a simultaneous (correlated) fit of ratio-correlators for all interpolators and momentum smearing choice to the model function $R_{p,z}^{\Gamma,i}(t) = M_{p,z} + A \exp(-\Delta E t)$. In the usual Bayesian fashion, $M$, $A$, and $\Delta E$ are specified by prior distributions. $M$, $A$, are drawn from normal distributions $\mathcal{N}(0,1)$ and $\mathcal{N}(0,0.1)$ respectively. A log-normal prior is assigned to excited state gap, $\ln(\Delta E) \sim \mathcal{N}(\overline{\ln(\Delta E_{\text{2pt}})}, \sigma=5\sigma[\ln(\Delta E_{\text{2pt}})])$, using the gap $\Delta E_{\text{2pt}}$ taken from fitting the matrix of interpolating two-point functions.
\begin{figure}[h!]
    \centering
     \includegraphics[scale=0.45]{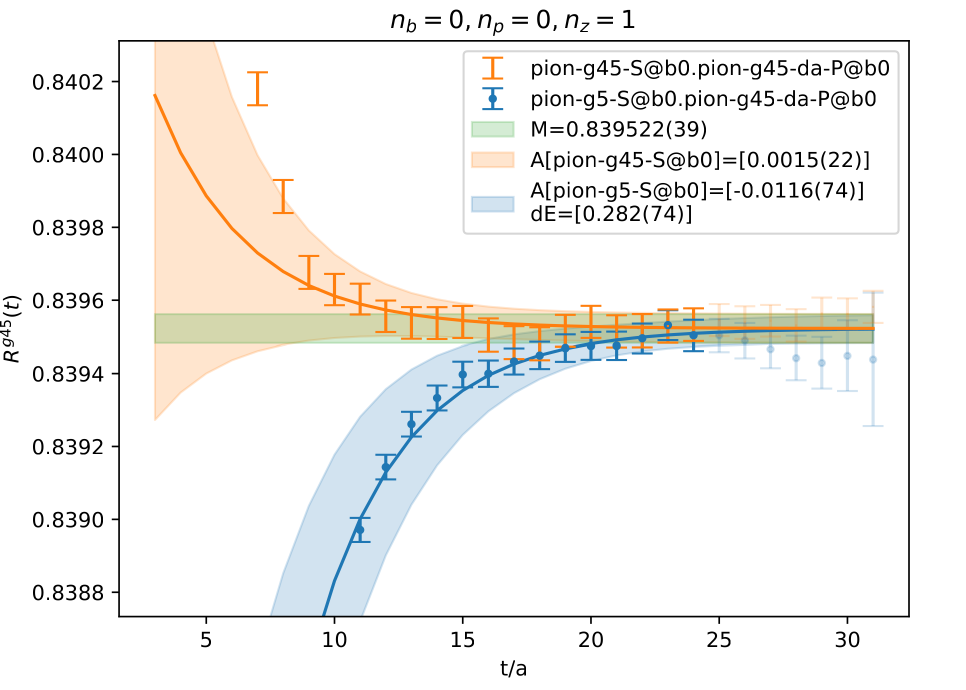}
    \includegraphics[scale=0.45]{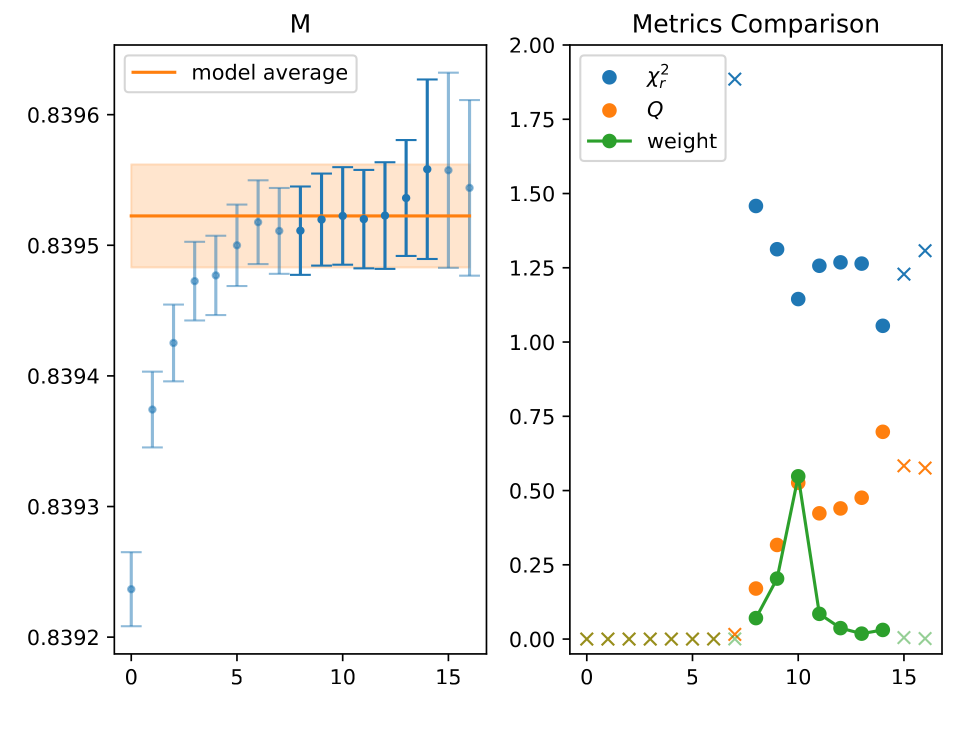}
    \includegraphics[scale=0.44]{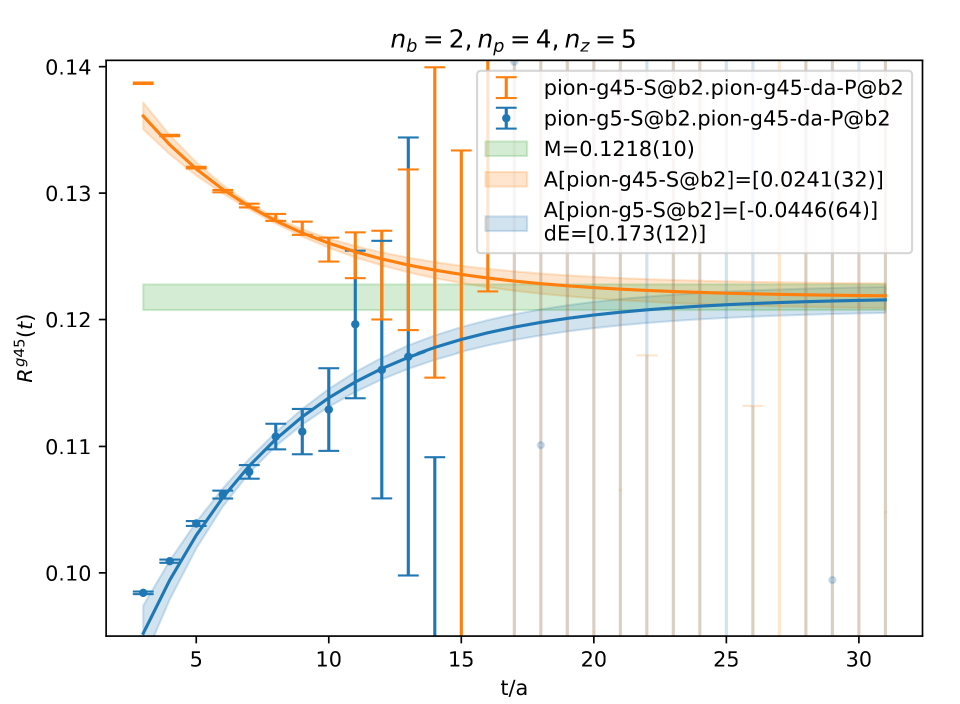}
    \includegraphics[scale=0.44]{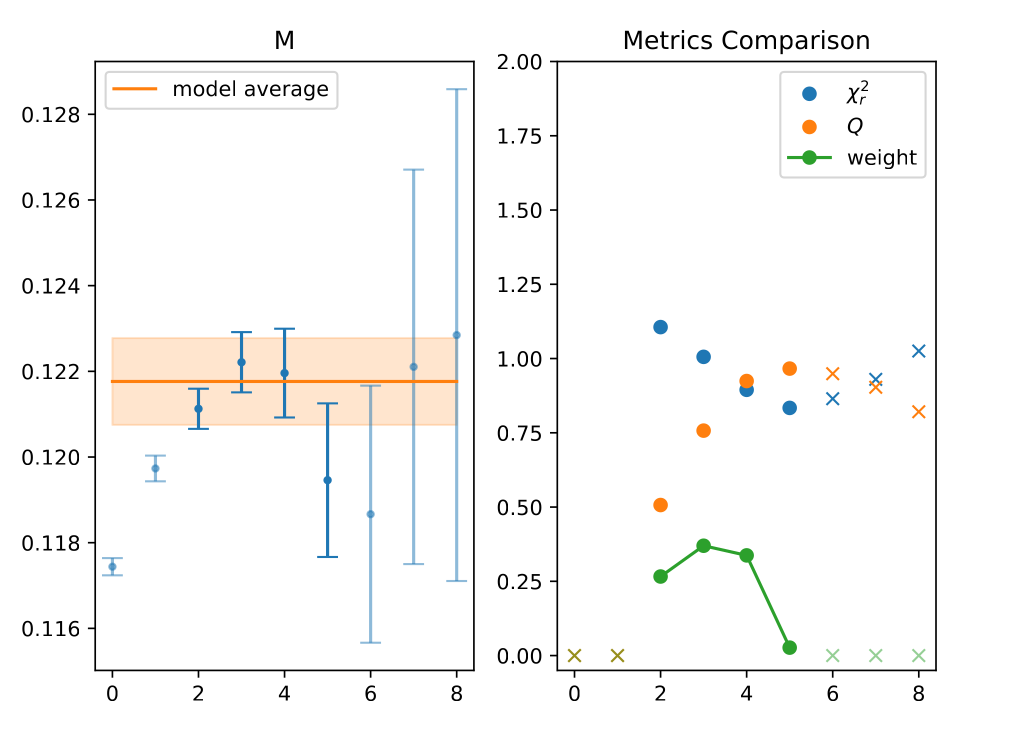}
    \caption{(first column) Model average fit against ratio-correlators at $p,z = 0,a$ and $p,z = 4\frac{2\pi}{La},5a$ (first, second row). (second column)  Profile of the matrix element as a function of number of $t_{\rm min}$ points removed. (third column) Various metric profiles as a function of number of $t_{\rm min}$ points removed.  The origin refers to the lowest $t_{\rm min}$ kept in the model average, $t_{\rm min} = 3a$. The lightly shaded points in the second two columns panels correspond to fits with normalized weights that are less than $1\%$, while the shaded points in the first column are excluded from the model average, but are plotted for convenience. }
    \label{fig:plots}
\end{figure}
We employ the Bayesian Model Averaging (BMA) framework proposed in ~\cite{Jay:2020jkz} to amalgamate uncertainties associated with the ability to resolve the excited states given two interpolators, and the underlying statistical uncertainties of the correlators. This is especially important at moderate-to-high momentum, where the exponentially decreasing signal-to-noise of pion correlators prevents fitting solely in the long Euclidean time regime. Analysis within BMA treats the choice of the fitting window as a particular model $M_i$ to the entire dataset $D$. In the range of the fitting window $t_{\rm keep}$, the data is modeled by a specific physical functional form (in the case of two-point functions, a sum of exponentials). Data outside the fitting window $t_{\rm cut}$ are modeled ``perfectly", i.e. the residual of the model and data in this region vanishes. Analytic marginalization of the posterior over these ``perfect" parameters incorporate the uncertainty in varying the fit window into the physically meaningful parameters $\theta$. 
This leads to the expression for the expectation value of quantity $f(\theta)$ weighted average over fit windowed quantities $\expval{f(\theta)}_i$:
\begin{align}
    \expval{f(\theta)} = \sum_{i} P(M_i | D)\expval{f(\theta)}_i,\quad \sum_i P(M_i | D) = 1.
\end{align}
For sufficiently large statistics, the weight is a modified Akaike Information Criterion (AIC) with parameter count $k_i$ modified by number of time slices outside the fit window $N_{{\rm cut},i}$, 
\begin{align}
    -2\log{P(M_i | D)} \propto \chi_{{\rm aug},i}^2 + 2 (k_i+N_{{\rm cut},i}),
\end{align}
where $\chi_{{\rm aug},i}^2$ includes both the correlated residual and prior of the model inside the fitting window.
In this analysis, the fitting window was changed by varying $t_{\rm min}$ to study the excited state contamination. The end of the fit window is fixed to be the time slice that is at least two thirds the time-like extent of the lattice whose signal-to-noise ratio is above some cutoff. Effects from finite temporal extent could have been studied by varying $t_{\rm max}$, but are unnecessary for current statistics.
 Representative BMA fits to the ratio of correlators, in Eq.~\eqref{eq:ratio}, are found in Fig.~\ref{fig:plots}, along with the profile of model weights and matrix element. 
 
 In the rest frame case (upper row of panels in Fig.~\ref{fig:plots}), each interpolator guarantees ground-state dominance in the long time regime, made evident by the constant signal-to-noise. One would expect that the maximally contributing fit ranges would be in the region where variations in $t_{\rm min}$ probe the strength of the contaminating exponentials, which seems to happen in the vicinity of $t_{\rm min}=13$, where the ratio-correlator with the $\gamma^5\gamma^t$ interpolator has reached a plateau and the ratio-correlator with $\gamma^5$ has not. This expectation is reflected in the model weight profile, where the maximally contributing fits have a $t_{\rm min}$ around this value.

 In the high-momentum case, the exponentially decreasing signal-to-noise makes ground-state dominance less clear, as seen in the second row of panels of Figure ~\ref{fig:plots}. The approach to the ground state region from both above and below still provides some constraint on the ground state plateau. This is demonstrated in the matrix element profile, where the individual fit results fluctuate in the interval bounded by both ratio-correlators. 

\begin{figure}[t!]
    \centering
    \includegraphics[scale=0.4]{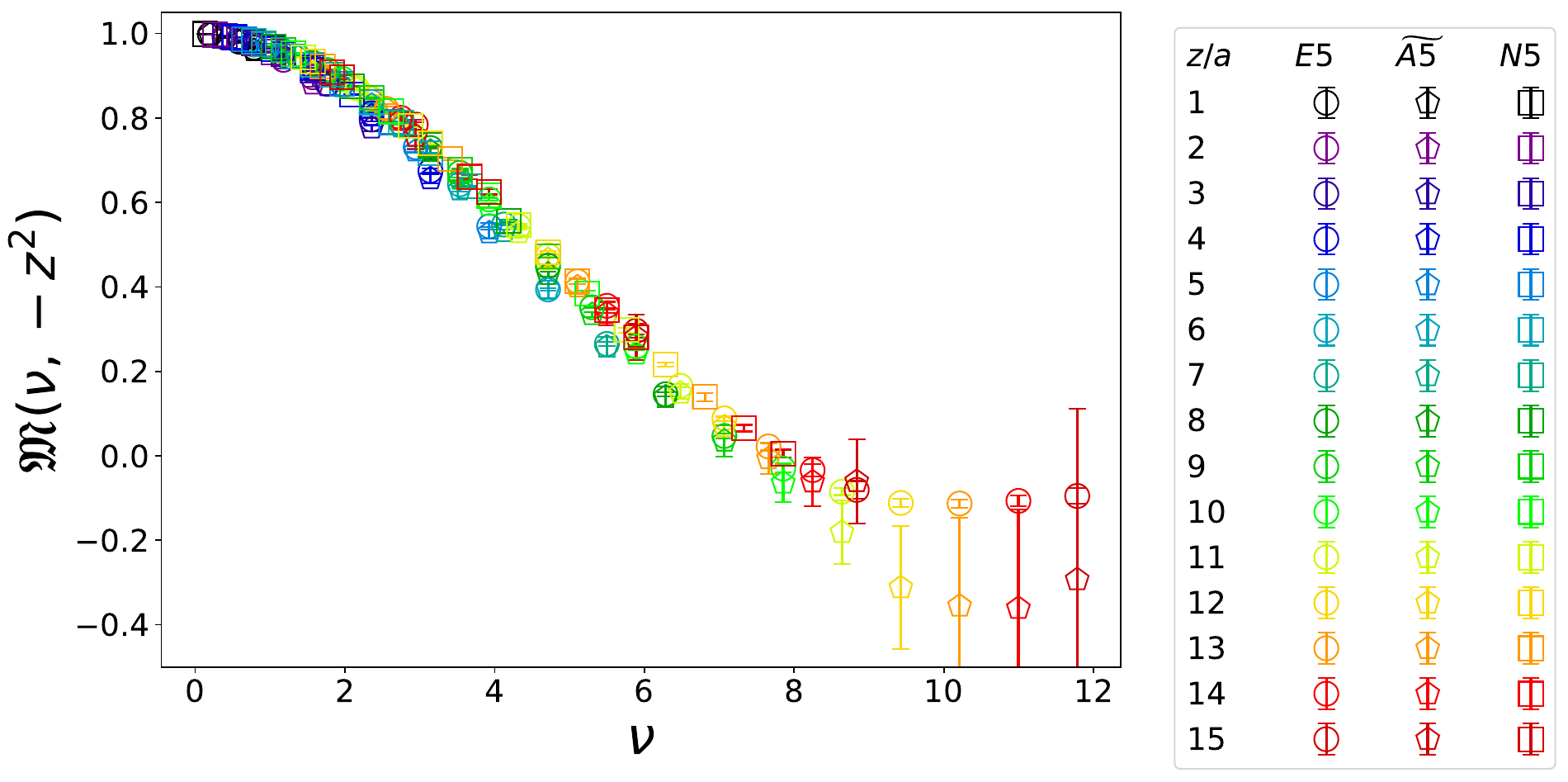}
    \caption{Reduced ITD $\mathfrak{M}(\nu,-z^2)$ for each ensemble listed in Table ~\ref{tab:lat}. }
    \label{fig:money}
\end{figure}

To propagate correlations between the matrix element at different kinematics, each fit is performed under jackknife to produce jackknife ensembles of model-averaged means $\{\expval{M_{p,z}}\}_J$ and variances $\{\sigma^2_{M_{p,z}}\}_J$. Ensemble $\{\expval{M_{p,z}}\}_J$ is rescaled such that the diagonal of its covariance matrix is the mean of $\{\sigma^2_{M_{p,z}}\}_J$. This rescaling preserves the correlation structure of the $\{\expval{M_{p,z}}\}_J$, while allowing the nonlinear error-propagation of $\{\sigma^2_{M_{p,z}}\}_J$. The reduced ITD, formed by taking the ratio of plateaus
\begin{align}
    \mathfrak{M}(\nu,-z^2) = \frac{M_{p,z}}{M_{0,z}} = \frac{M^\Gamma(p,z)/E(p)}{M^\Gamma(0,z)/E(0)} = \frac{\mathcal{M}(\nu,-z^2)}{\mathcal{M}(0,-z^2)},
\end{align}
is shown in Figure~\ref{fig:money}.

\section{Conclusion/Future Prospects}

We have presented progress on the lattice calculation of the LCDA, resulting in a reduced ITD which may be used to extract the LCDA through perturbative matching. The reduced ITD has been computed at three different lattice spacings, at a pion mass of $\sim440$ MeV.  Proceeding with the LCDA extraction requires a remedy to the inverse problem associated with the inversion of convolution integrals of $\mathfrak{M}$ over all $\nu$, of which the lattice only provides a finite sampling of $\mathfrak{M}$ over a finite range of $\nu$. Several different approaches to this inverse problem have been explored in~\cite{Karpie:2019eiq,Alexandrou:2020tqq,DelDebbio:2020rgv,Karpie:2021pap}. A popular approach involves fitting several models of the LCDA to the data, exchanging the inverse problem for a model-dependency problem. Lattice-spacing, pion-mass, off-lightcone, and other effects may be modeled similarly, provided that $z^2$, $a$, etc are sufficiently small. The percentage of the data that satisfies these constraints is not known a priori. BMA naturally lends itself to this sort of problem and will be crucial to the rest of the analysis. Extending the currently employed ensemble set to smaller pion masses is underway.
how 

{\bf Acknowledgements:}
We are grateful to the ALPHA collaboration and the CLS effort for sharing some of their ensembles with us.
This work was supported by the U.S. DOE Grant \#DE-FG02-04ER41302 and by the US DOE Contract No. \#DE-AC05-06OR23177. 
A. R. acknowledges support by the Jefferson Science Associates,
 LLC under  U.S. DOE Contract \mbox{ \#DE-AC05-06OR23177} 
 and by U.S. DOE Grant \#DE-FG02-97ER41028.  
S.Z.~acknowledges support by the French Centre national de la recherche scientifique (CNRS) under an Emergence@INP 2023 project. 
We would also like to thank the Texas Advanced Computing Center (TACC) at the University of Texas at Austin for providing HPC resources
on Frontera~\cite{frontera} that have contributed to the results in this paper. 
This work also benefited from access to the Jean Zay supercomputer at the Institute for Development and Resources in Intensive Scientific Computing (IDRIS) in Orsay, France under project 2022-A0080511504. 
 In addition, this work used resources at NERSC, a DOE Office of Science User Facility supported by the Office of Science of the U.S. Department of Energy under Contract \#DE-AC02-05CH11231, as well as resources of the Oak Ridge Leadership Computing Facility at the Oak Ridge National Laboratory, which is supported by the Office of Science of the U.S. Department of Energy under Contract No. \mbox{\#DE-AC05-00OR22725}.

\bibliographystyle{apsrev4-2} 
\bibliography{biblio}

\end{document}